\documentclass[journal]{IEEEtran}

\ifCLASSINFOpdf
\else
   \usepackage[dvips]{graphicx}
\fi
\usepackage{url}

\usepackage{cite}
\usepackage{amsmath,amssymb,amsfonts}
\usepackage{algorithmic}
\usepackage{graphicx}
\usepackage{textcomp}
\usepackage{xcolor}
\usepackage{bm}
\usepackage{booktabs} 
\usepackage{multirow}

\usepackage{array}

\newcommand{\bb}[1]{\mathbf{#1}}

\begin{document}

\title{Optimal Sensor Placement for TDOA-Based Source Localization with Sensor Location Errors}

\author{Chengjie Zhang, Xinyang Han \thanks{Emails: 12332644@mail.sustech.edu.cn}}

\markboth{Journal of \LaTeX\ Class Files, Vol. 14, No. 8, August 2024}
{Shell \MakeLowercase{\textit{et al.}}: Bare Demo of IEEEtran.cls for IEEE Journals}
\maketitle

\begin{abstract}



The accuracy of time difference of arrival (TDOA)-based source localization is influenced by sensor location deployment. Many studies focus on optimal sensor placement (OSP) for TDOA-based localization without sensor location noises (OSP-WSLN). In practice, there are sensor location errors due to installation deviations, etc, which implies the necessity of studying OSP under sensor location noises (OSP-SLN). There are two fundamental problems: What is the OSP-SLN strategy? To what extent do sensor location errors affect the performance of OSP-SLN? For the first one, under the assumption of the near-field and full set of TDOA, minimizing the trace of the Cramér–Rao bound is used as optimization criteria. Based on this, a concise equality, namely Eq. (\ref{eq:linearEq}), is proven to show that OSP-SLN is equivalent to OSP-WSLN. Extensive simulations validate both equality and equivalence and respond to the second problem: not large sensor position errors give an ignorable negative impact on the performance of OSP-SLN quantified by the trace of CRB. Also, simulations show source localization accuracy with OSP-SLN outperforms that with random placement. These simulations validate our derived OSP-SLN and its effectiveness. We have open-sourced the code for community use.

\end{abstract}


\IEEEpeerreviewmaketitle

\section{Introduction}
TDOA-based source localization is one of the very important localization technologies with a wide range of applications in navigation, search/rescue tasks, and mobile communications \cite{uav-nav, 2005positioning, 2011positioning}. In the past few decades, relevant research has mainly focused on designing estimation methods to improve the accuracy of TDOA-based source localization \cite{tdoa-loc1, tdoa-loc2, tdoa-loc3}, with less attention paid to the impact of sensor position configuration on source localization. In fact, the location of the sensor significantly affects the accuracy of source localization \cite{noiseeff}. Poor sensor placement can lead to poor localization results, and even cause localization failure. 

\begin{figure}[htbp]
\centerline{\includegraphics[width = 1.0\linewidth]{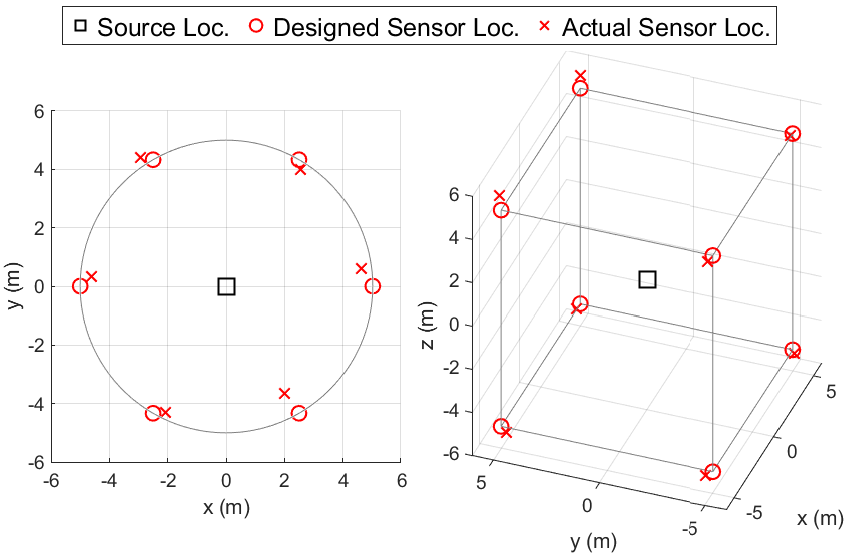}}
\caption{2D (left side) and 3D  (right side) cases of placing sensors following a designed strategy (gray trajectory) occurring deploying errors. `Designed Sensor Loc.'/`Actual Sensor Loc.' means deploying sensors following the designed strategy without/with sensor locations `Sound Loc.' is the true value of source location.}
\label{fig:SLNcase}
\end{figure}

Early works utilized Cramér-Rao bound (CRB) or Fisher information matrix (FIM) to find optimal sensor placement (OSP) strategies for TDOA-based source localization. \cite{2005trace} minimizes the trace of CRB for source position parameters to derive the necessary and sufficient conditions for the 2D/3D optimal sensor array configuration with assumptions of near-field and using full-set TDOA defined in Eq. (\ref{eq:tdoa}). \cite{2006trace} further analyzed the sensor placement strategy to improve TDOA-based source localization in other cases not considered in \cite{2005trace}. \cite{2007trace} studied different optimization criteria and derived some new optimal geometric configurations. \cite{2007fisher} derived the necessary and sufficient conditions for the optimal array configuration by maximizing the determinant of FIM and \cite{2009fisher} gave solutions to the OSP problem for an uncertain source location by maximizing the expected value of FIM's determinant.

Recently, \cite{hybrid-OSP1} and \cite{hybrid-OSP2} derived the OSP strategy based on fused TOA-RSS-AOA and hybrid TDOA-TOA-RSS-AOA respectively. \cite{unifiedADMM} have formulated a general OSP problem for various measurements including TOA, TDOA,
AOA or RSS and adopted a unified optimization-based framework to solve it.

In practice, deploying sensor placement strategies, including optimal sensor placement strategies, is inevitable to encounter sensor position errors due to installation deviations or changes in sensor location caused by long-term use, etc. Fig. \ref{fig:SLNcase} presents 2D and 3D examples of sensor placement following a specified strategy (gray trajectory) with sensor location errors. `Designed Sensor Loc.' can also be understood as the sensor locations we know, which are also regarded as the sensor location measurements. However, most works were related to optimal sensor placement without sensor location noises (OSP-WSLN), and only a few focused on optimal sensor placement with sensor location noises (OSP-SLN). \cite{2021-OSP-SLN} studied the influence of sensor location errors on the trace of CRB for TDOA-based localization. However, they only proved that the trace of CRB under sensor location errors is always larger than that of CRB without the errors and did not derive a solution to minimize the trace of CRB under sensor location errors, i.e., OSP-SLN. Very recently, in cooperative multi-robot localization, \cite{uav-sln-osp} studied optimal anchor (robot) placements under mobile anchor location uncertainty. They utilized range-based measurements which are different from ours.
There are three main contributions to this paper:
\begin{itemize}
    \item To our best knowledge, it is the first time exploring OSP-SLN for TDOA-based localization. Minimizing the trace of CRB is used to find out the OSP-SLN strategy.
    \item For the full set of TDOA defined in Eq. (\ref{eq:tdoa}) and scenario of near-field, we prove concise equality i.e., Eq. (\ref{eq:linearEq}) indicating that OSP-SLN is equivalent to OSP-WSLN.
    \item We conducted simulations to show the correctness of the equality, the equivalence between OSP-WSLN and OSP-SLN, no large sensor location noises taking the ignorable influence on the performance of OSP-SLN quantified by the trace of CRB, and the contribution of OSP-SLN to improving source localization accuracy. We open-source our code to benefit the community.
\end{itemize}

\section{Formulation}
\label{sec:format}
Denote the $i$-th sensor location as $\mathbf{x}_i\in\mathbb{R}^D$ ($i=1,2,...,N$) and $\mathbf{s}\in\mathbb{R}^D$ is the location of the sound source ($D=2,3$). Denote measurement and measurement function for time difference of arrival (TDOA) between the $i$-th sensor and the $j$-th sensor are $t_{i,j}$ and $T_{i,j}$ respectively and their relationship is
\begin{equation}
\label{eq:tdoa}
t_{i,j}=T_{i,j}+w_{i,j}=\frac{||\bb{s}-\bb{x}_i||  -  ||\bb{s}-\bb{x}_j||}{c} + w_{i,j}
\end{equation}
where $c$ is the propagation velocity of source, and $w_{i,j}\sim N(0, \sigma_t^2)$ is Gaussian noise in $t_{i,j}$. Denote TDOA measurements as $\bb{t}=[t_{2,1},t_{3,1},...,t_{N,1},t_{3,2},...,t_{N,N-1}]^T\in\mathbb{R}^M$ ($M=\frac{N(N-1)}{2}$) and the corresponding vector of measurement function $\bb{T}$ also arrange its elements in this order.
Denote measurement and measurement function of the $i$-th sensor location are $\bb{m}_i$ and $\bb{x}_i$ respectively and their relationship is
\begin{equation}
\label{eq:sensor}
\bb{m}_i=\bb{x}_i + \bb{v}_i
\end{equation}
where $\bb{v}_i\sim N(\bb{0}, \sigma_{loc}^2\bb{I}_D)$ that is Gaussian noise. Denote sensor location measurement as $\bb{m}=[\bb{m}_1,\bb{m}_2,...,\bb{m}_N]^T\in\mathbb{R}^{DN}$ and the corresponding vector of measurement function $\bb{x}$ also arrange its elements in this order.

Denote $\boldsymbol{\theta}$ as the parameter vector that needs to be estimated and $\bb{z}$ is the measurement vector. When sensor locations are precisely known, $\bb{z}=\bb{t}$, $\boldsymbol{\theta}=\bb{s}$, $\bb{g}(\boldsymbol{\theta})=\bb{T}$ and when sensor locations have errors, $\bb{z}=[\bb{t},\bb{m}]^T$ $\boldsymbol{\theta}=[\bb{s},\bb{x}]^T$, $\bb{g}(\boldsymbol{\theta})=[\bb{T},\bb{x}]^T$. The maximum likelihood function is formulated as
\begin{equation}
\label{eq:ml}
L(\boldsymbol{\theta})=G exp[-\frac{1}{2}(\bb{g}(\boldsymbol{\theta})-\bb{z})^T\bb{\Sigma}^{-1}(\bb{g}(\boldsymbol{\theta})-\bb{z})]
\end{equation}
where $\bb{\Sigma}$ is the covariance matrix with diagonal form and $G$ is a constant. The Fisher information matrix $\bb{F}$ is 
\begin{equation}
\label{eq:fisherMatix}
    \bb{F}=E[[\triangledown_{\boldsymbol{\theta}} ln L(\boldsymbol{\theta})][\triangledown_{\boldsymbol{\theta}} ln L(\boldsymbol{\theta})]^T] _{\boldsymbol{\theta}=\boldsymbol{\theta}_0}=\bb{J}^T\bb{\Sigma}^{-1}\bb{J}
\end{equation}
where $\boldsymbol{\theta}_0$ is the true value of $\boldsymbol{\theta}$. The Jacobian matrix and the CRB matrix are $\bb{J}=\frac{\partial \bb{g}(\boldsymbol{\theta})}{\partial \boldsymbol{\theta}}$ and $\bb{C}=\bb{F}^{-1}$ respectively. Denote the CRB matrix and the Jacobian matrix with/without sensor location errors as $\bb{C}$/$\bb{C}'$ and $\bb{J}$/$\bb{J}'$ respectively. Denote $\bb{J}$ and $\bb{J}'$ as
\begin{equation}
\label{eq:J}
\begin{aligned}
\bb{J} &=
\begin{bmatrix}
\frac{\partial \bb{T}}{\partial \bb{s}} & \frac{\partial \bb{T}}{\partial \bb{x}} \\
\frac{\partial \bb{x}}{\partial \bb{s}} & \frac{\partial \bb{x}}{\partial \bb{x}}
\end{bmatrix}
=
\begin{bmatrix}
\bb{J}_1 & \bb{J}_2\\
\bb{0} & \bb{J}_3
\end{bmatrix}, \bb{J}'=\bb{J}_1
\end{aligned}
\end{equation}
respectively where $\bb{J}_3=\bb{I}_{DN}$.

\section{Equivalence between OSP-SLN and OSP-WSLN}
In this section, we first derive the concise equality describing the linear relationship between the trace of CRB for OSP-SLN and the trace of CRB for OSP-WSLN. Then, the equivalence between OSP-SLN and OSP-WSLN can be easily proven based on the equality Eq. (\ref{eq:linearEq}).

Based on (\ref{eq:fisherMatix}) and (\ref{eq:J}), CRB matrix under sensor location errors $\bb{C}$ is
\begin{equation}
\label{eq:C}
\bb{C} =
(\begin{bmatrix}
\bb{J}_1^T & \bb{0} \\
\bb{J}_2^T & \bb{J}_3^T
\end{bmatrix}
\begin{bmatrix}
\sigma_t^{-2} \bb{I}_M & \bb{0} \\
\bb{0} & \sigma_{loc}^{-2} \bb{I}_{ND}
\end{bmatrix}
\begin{bmatrix}
\bb{J}_1 & \bb{J}_2\\
\bb{0} & \bb{J}_3
\end{bmatrix})^{-1}.
\end{equation}

Denote $\bb{C}_1$ as the sub-matrix of $\bb{C}$, which represents the CRB matrix for sound location parameters. By the block matrix inversion formula and simplification, we have
\begin{equation}
\label{eq:C1}
\begin{aligned}
\bb{C}_1 =
&\sigma_t^2[(\bb{J}_1^T \bb{J}_1)^{-1} + (\bb{J}_1^T \bb{J}_1)^{-1} \bb{J}_1^T \bb{J}_2 (\bb{J}_2^T \bb{J}_2 - \bb{J}_2^T \bb{J}_1
\\
&(\bb{J}_1^T \bb{J}_1)^{-1} \bb{J}_1^T \bb{J}_2 + \frac{\sigma_t^2}{\sigma_{loc}^2} \bb{J}_3^T \bb{J}_3)^{-1} \bb{J}_2^T \bb{J}_1 (\bb{J}_1^T \bb{J}_1)^{-1}].
\end{aligned}
\end{equation}

Here, $\bb{C}'=\sigma_t^2(\bb{J}_1^T \bb{J}_1)^{-1}$. The trace of $\bb{C}_1$ i.e., $tr\{\bb{C}_1\}$ and the trace of $\bb{C}'$ i.e., $tr\{\bb{C}'\}$ represent the lower bound for the mean square error (MSE) of the source location with and without sensor location noises respectively. To further analyze, we expand Eq. (\ref{eq:C1}) by doing the singular value decomposition (SVD) of $\bb{J}_1$:
\begin{equation}
\label{eq:SVD_J1}
\bb{J}_1 = \bb{U}_1 \bb{\Sigma}_1 \bb{V}_1^T,
\end{equation}
where $\bb{U}_1\in\mathbb{R}^{M\times D}$ and unitary matrix $\bb{V}_1\in\mathbb{R}^{D\times D}$. Then we have
\begin{equation}
\label{eq:J1term}
\bb{J}_1 (\bb{J}_1^T \bb{J}_1)^{-K} \bb{J}_1^T = \bb{U}_1 (\bb{\Sigma}_1^T \bb{\Sigma}_1)^{-K+1} \bb{U}_1^T
\end{equation}
where $K=1,2$ and $(\bb{\Sigma}_1^T \bb{\Sigma}_1)^0=\bb{I}_D$. Based on Eq. (\ref{eq:J1term}), we simplify the trace of $\bb{C}_1$:
\begin{equation}
\label{eq:simplified trace of C1}
\begin{aligned}
tr\{\bb{C}_1\} &= tr\{\bb{C}'\} + \sigma_t^2tr\{\bb{U}_1 (\bb{\Sigma}_1^T \bb{\Sigma}_1)^{-1} \bb{U}_1^T
\\
&\bb{J}_2 [\bb{J}_2^T (\bb{I}_M - \bb{U}_1\bb{U}_1^T) \bb{J}_2 + \frac{\sigma_t^2}{\sigma_{loc}^2} \bb{I}_{DN}]^{-1} \bb{J}_2^T\}.
\end{aligned}
\end{equation}

Furthermore, a gap function $f(\bb{s},\bb{x},\sigma_t,\sigma_{loc})$ between $tr\{\bb{C}_1\}$ and $tr\{\bb{C}'\}$ is
\begin{equation}
\label{eq:gap}
\begin{aligned}
&f(\bb{s},\bb{x},\sigma_t,\sigma_{loc}) = tr\{\bb{C}_1\} - tr\{\bb{C}'\} = \sigma_t^2tr\{(\bb{\Sigma}_1^T \bb{\Sigma}_1)^{-1} \\& \underbrace{\bb{U}_1^T\bb{J}_2 [\bb{J}_2^T (\bb{I}_M - \bb{U}_1\bb{U}_1^T) \bb{J}_2 + \frac{\sigma_t^2}{\sigma_{loc}^2} \bb{I}_{DN}]^{-1} \bb{J}_2^T\bb{U}_1}_{\bb{L}}\}.
\end{aligned}
\end{equation}

Unless otherwise specified, $f$ is used to stand for $f(\bb{s},\bb{x},\sigma_t,\sigma_{loc})$. Next, we do the SVD on $\bb{J}_2$:
\begin{equation}
    \label{eq:svd_J2}
     \bb{J}_2 = \bb{U}_2 \bb{\Sigma}_2 \bb{V}_2^T 
\end{equation}
where $\bb{U}_2\in\mathbb{R}^{M\times N-1}$ and unitary matrix $\bb{V}_2\in\mathbb{R}^{DN\times DN}$. Based on Eq. (\ref{eq:svd_J2}), $\bb{L}$ is simplified as
\begin{equation}
    \label{eq:sim_L}
    \begin{aligned}
\bb{L}=\bb{U}^T_1\bb{U}_2\bb{\Sigma}_2&[\bb{\Sigma}^T_2\bb{U}^T_2(\bb{I}_M - \bb{U}_1\bb{U}_1^T)\bb{U}_2\bb{\Sigma}_2\\&+\frac{\sigma_t^2}{\sigma_{loc}^2} \bb{I}_{DN}]^{-1}\bb{\Sigma}^T_2\bb{U}^T_2\bb{U}_1
    \end{aligned}
\end{equation}
where $\bb{\Sigma}_2 = \begin{bmatrix}\bb{D}_2 & \bb{0}\end{bmatrix}$. By Lemma 1 demonstrated at the end of this section, $\bb{D}_2=\frac{\sqrt{N}}{c}\bb{I}_{N-1}$. $\bb{L}$ is further simplified as
\begin{equation}
    \label{eq:sim_L2}
    \begin{aligned}
\bb{L}=\bb{U}^T_1\bb{U}_2&[\bb{U}^T_2(\bb{I}_M - \bb{U}_1\bb{U}_1^T)\bb{U}_2\\&+\frac{1}{NK} \bb{I}_{N-1}]^{-1}\bb{U}^T_2\bb{U}_1
    \end{aligned}
\end{equation} where
\begin{equation}
    K=\frac{\sigma^2_{loc}}{c^2\sigma^2_t}.
\end{equation}

Assume $\bb{A}=\bb{U}^T_2\bb{U}_1$ and according to Lemma 2, the SVD is performed: $\bb{A}=\bb{U}_A\bb{\Sigma}_A\bb{V}^T_A$ where $\bb{\Sigma}_A=\begin{bmatrix}
    \bb{I}_D \\ \bb{0}
\end{bmatrix}\in\mathbb{R}^{N-1\times D}$. Note that Lemma 2 is proven at the end of this section. Both $\bb{U}_A$ and $\bb{V}_A$ are unitary matrix. Note that $\bb{U}^T_2\bb{U}_2=\bb{I}$ and replace $\bb{U}^T_2\bb{U}_1$ with $\bb{A}$, we get
\begin{equation}
\label{eq:sim_L4}
\begin{aligned}
    \bb{L}=&\bb{A}^T[-\bb{A}\bb{A}^T+(1+\frac{1}{NK})\bb{I}_{N-1}]^{-1}\bb{A} \\
=&\bb{V}_A\bb{\Sigma}_A^T(\bb{\Sigma}_A\bb{\Sigma}_A^T+\frac{1}{NK} \bb{I}_{N-1})^{-1}\bb{\Sigma}_A\bb{V}_A^T \\
=&NK\bb{I}_D.
\end{aligned}
\end{equation}
Therefore, \begin{equation}
\label{eq:gap3}
\begin{aligned}
    f&=\sigma^2_t tr\{(\bb{\Sigma}_1^T \bb{\Sigma}_1)^{-1}\bb{L}\}\\
&=NK tr\{\bb{C}'\},
\end{aligned}
\end{equation}
which means 
\begin{equation}
\label{eq:linearEq}
    tr\{\bb{C}_1\}=(1+NK)tr\{\bb{C}'\}.
\end{equation}

Therefore, without loss of generality, assume $\bb{s}=\bb{0}$, it is easy to prove the equivalence:  $\underset{\bb{x}}{\min} \ tr\{\bb{C}_1\} 
\Leftrightarrow \underset{\bb{x}}{\min} \ tr\{\bb{C}'\}$, which indicates that OSP-WSLN is equivalent to OSP-SLN.

\textbf{Lemma 1:} When $\bb{J}_2$ is decomposed by the thin SVD i.e., Eq. (\ref{eq:svd_J2}), $\bb{\Sigma}_2 = \begin{bmatrix}\bb{D}_2 & \bb{0}\end{bmatrix}$ and $\bb{D}_2=\frac{\sqrt{N}}{c}\bb{I}_{N-1}$. 

\textit{Proof:} $\bb{J}_2$ is formulated as \begin{equation}
    \label{eq:J2}
    \begin{aligned}
    \bb{J}_2 &= \frac{1}{c}\begin{bmatrix}
        \bb{u}_1^T & -\bb{u}_2^T & \bb{0} & \cdots & \bb{0} &  \bb{0}\\
        \bb{u}_1^T & 0 & -\bb{u}_3^T & \cdots & \bb{0} & \bb{0}\\
        \vdots & \vdots & \vdots & \ddots & \vdots & \vdots\\
        \bb{u}_1^T &  \bb{0} &  \bb{0} & \cdots & \bb{0} & -\bb{u}_N^T\\
        \bb{0} & \bb{u}_2^T & -\bb{u}_3^T &  \cdots & \bb{0} &  \bb{0}\\
        \vdots & \vdots & \vdots & \ddots & \vdots & \vdots\\
        \bb{0} & \bb{u}_2^T & \bb{0} &  \cdots & -\bb{u}_{N-1}^T &  \bb{0}\\
        \bb{0} & \bb{u}_2^T &  \bb{0} & \cdots & \bb{0} & -\bb{u}_N^T\\
        \vdots & \vdots & \vdots & \vdots & \vdots & \vdots\\
        \bb{0} & \bb{0} & \bb{0} &  \cdots & -\bb{u}_{N-1}^T &  \bb{u}_N^T
        \end{bmatrix}
    \end{aligned}
\end{equation} where $\bb{u}_i^T=\frac{(\bb{s}-\bb{x}_i)^T}{  ||\bb{s}-\bb{x}_i||}$ is a unit row vector. Denote symmetric $\bb{B}=\bb{J}_2\bb{J}^T_2$ and $\bb{B}$ is full rank. The formulation of $\bb{B}$ is ignored due to the limited space. An important property of $\bb{B}$ is: $\bb{B}^T\bb{B}=\frac{N}{c^2}\bb{B}$, which can be demonstrated by calculation. Therefore, assume $\lambda$ and $\bb{q}$ are any eigenvalue and the corresponding eigenvector of $\bb{B}$ respectively, $\bb{B}\bb{q}=\lambda\bb{q}\Longrightarrow\bb{B}^T\bb{B}\bb{q}=\lambda\bb{B}^T\bb{q}\Longrightarrow\frac{N}{c^2}\bb{q}=\lambda\bb{q}$. Because any eigenvector of $\bb{B}$ of full rank is non-zero, $\lambda=\frac{N}{c^2}$. Therefore, non-zero singular values of $\bb{J}_2$ equal to $\frac{\sqrt{N}}{c}$, which means $\bb{D}_2=\frac{\sqrt{N}}{c}\bb{I}_{N-1}$.

\textbf{Lemma 2:} When we do the SVD on $\bb{A}$ ($\bb{A}=\bb{U}_2^T\bb{U}_1$): $\bb{A}=\bb{U}_A\bb{\Sigma}_A\bb{V}^T_A$, $\bb{\Sigma}_A=\begin{bmatrix}
    \bb{I}_D \\ \bb{0}
\end{bmatrix}\in\mathbb{R}^{N-1\times D}$.

\textit{Proof:} we formulate $\bb{J}_1$ as
 \begin{equation}
    \label{eq:J1}
    \begin{aligned}
    \bb{J}_1 = \frac{1}{c}
    \begin{bmatrix}
        \bb{J}_{1,1}\\
        \bb{J}_{1,2}\\
        \vdots \\
        \bb{J}_{1,N-1}\\
    \end{bmatrix} \text{where} \
        \bb{J}_{1,i} = \begin{bmatrix}
        \bb{u}_{i+1}^T-\bb{u}_i^T \\
        \bb{u}_{i+2}^T-\bb{u}_i^T \\
        \vdots \\
        \bb{u}_N^T-\bb{u}_i^T \\
        \end{bmatrix}_{(N-i)\times D}.
    \end{aligned}
\end{equation}

Note that $\bb{J}_2$ is full column rank and each column vector in $\bb{J}_1$ is the linear combination of column vectors in $\bb{J}_2$. The latter can be formulated as $\bb{J}_1=\bb{J}_2\bb{X}$. Using both Eq. (\ref{eq:SVD_J1}) and Eq. (\ref{eq:svd_J2}), $\bb{J}_2\bb{X}=\bb{U}_1 \bb{\Sigma}_1 \bb{V}_1^T\Longrightarrow \bb{U}_2 \bb{\Sigma}_2 \bb{V}_2^T\bb{X}=\bb{U}_1 \bb{\Sigma}_1 \bb{V}_1^T\Longrightarrow\bb{U}_2 \underbrace{\bb{\Sigma}_2 \bb{V}_2^T\bb{X}\bb{V}_1\bb{\Sigma}_1^{-1}}_{\bb{Y}}=\bb{U}_1\Longrightarrow\bb{U}_2\bb{Y}=\bb{U}_1$. 

Because $\bb{U}_1^T\bb{U}_1=\bb{I}$ and $\bb{U}_2^T\bb{U}_2=\bb{I}$, $\bb{U}_1^T\bb{U}_1=\bb{Y}^T\bb{U}^T_2\\\bb{U}_2\bb{Y}=\bb{Y}^T\bb{Y}=\bb{I}$. Then, $\bb{A}^T\bb{A}=\bb{Y}^T\bb{U}^T_2\bb{U}_2\bb{U}_2^T\bb{U}_2\bb{Y}=\bb{I}\Longrightarrow \bb{\Sigma}_A^T\bb{\Sigma}_A=\bb{I}\Longrightarrow \bb{\Sigma}_A=\begin{bmatrix} \bb{I}_D \\ \bb{0}\end{bmatrix}$.


\section{Simulations}
We need to verify four parts: demonstrating the equality Eq. (\ref{eq:linearEq}), the equivalence between OSP-SLN and OSP-WSLN, the tiny impact of not large sensor location errors on the performance of OSP-SLN, and the superiority of OSP-SLN in source localization. Indoor sound source localization is used as an example.
\subsection{Correctness of the Equality}
To fully verify the correctness of the derived equality, we randomly generate all parameters except for the sound velocity within a certain range: $D=2,3, N=4,5,6,..., 20, 10^{-5}s  \leq \sigma_t \leq 10^{-3}s, 0.01m  \leq \sigma_{loc} \leq 1m$. The source is set at the origin and the sensor positions are randomly generated in a square space with a side length of $10m$. 100,000 Monte-Carlo simulations are performed. In each Monte-Carlo run, we record $N$, $K$, $tr\{\bb{C}_1\}$, and $tr\{\bb{C}'\}$ and calculate $err=|tr\{\bb{C}_1\}-(1+NK)tr\{\bb{C}'\}|$. After 100,000 Monte-Carlo simulations, the mean of $err$ equals $6.44\times10^{-15}$ after removing outliers in boxplot \cite{modernStatic}. The averaged error, i.e., $6.44\times10^{-15}$ is inevitably caused by floating-point calculations, which indicates the concise equality Eq. (\ref{eq:linearEq}) holds when ignoring numerical errors.

\begin{figure}[htbp]
\centerline{\includegraphics[width = 0.93\linewidth]{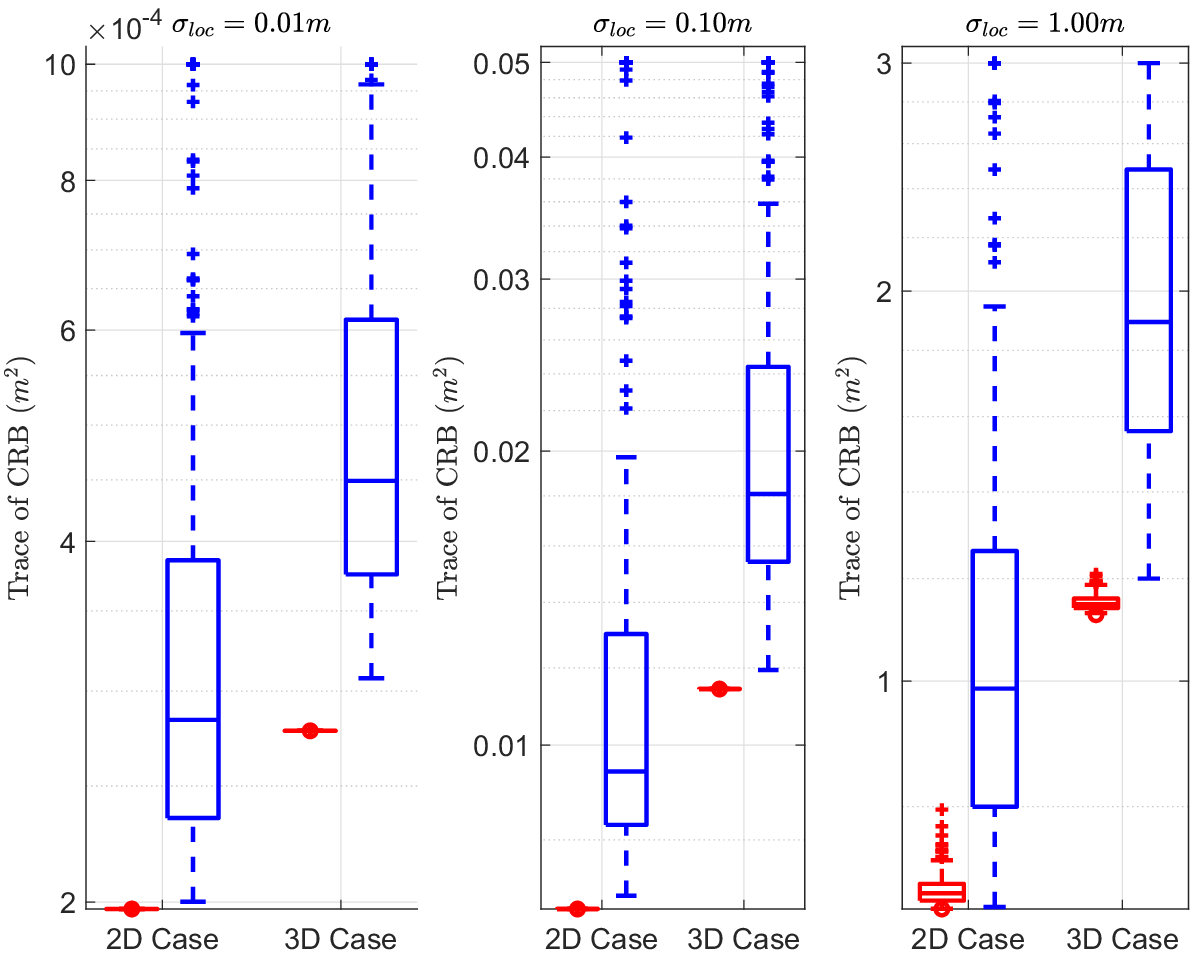}}
\caption{Simulation results of $tr\{\bb{C}^o_1\}$ (red circle), $tr\{\bb{C}^s_1\}$ (red box plot) and $tr\{\bb{C}^{ran}_1\}$ (blue box plot) in 2D/3D space under three levels of sensor location noises ($\sigma_{loc}=0.001m,0.01m,0.1m$).}
\label{fig:exp2}
\end{figure}

\subsection{Verification of the Equivalence and Minute Impact}
\subsubsection{Set up}
In this section, two conclusions are to be verified: the equivalence between OSP-SLN and OSP-WSLN, and the tiny impact of limited sensor location errors on the performance of deploying the OSP-SLN strategy. Under assumptions of the full set of TDOA and near-field, the OSP-WSLN in 2D space is a uniform angular array (UAA) and the OSP-WSLN in 3D space is a platonic solid \cite{2005trace}. UAA is set as
\begin{equation}
    \begin{aligned}
        \alpha_i&=\alpha_0+\frac{2\pi}{N}(i-1),\\
        \bb{x}_i&=[r cos(\alpha_i), r sin(\alpha_i)]^T
    \end{aligned}
\end{equation}
where we set $\alpha_0=0$, $r=5m$, $i=1,2,...,N$, $N=6$, $D=2$. For platonic solids in 3D space, we select the cube whose $N=8$ and edge length equals $10m$. The source is placed at the origin. We set $\sigma_t=0.1ms$ and $\sigma_{loc}=0.01m, 0.1m, 1m$. $\bb{C}_1$ is called as $\bb{C}^o_1$ when true values of sensor location $\bb{x}$ precisely follow the optimal configuration. $\bb{C}_1$ is called as $\bb{C}^s_1$ when $\bb{x}$ equals optimal locations adding errors. $\bb{C}_1$ is called as $\bb{C}^{ran}_1$ when $\bb{x}$ are generated randomly over a square space with a side length of $10m$. In each $\sigma_{loc}$, 200 Monte-Carlo simulations are performed and in each simulation, $tr\{\bb{C}^o_1\}$, $tr\{\bb{C}^s_1\}$ and $tr\{\bb{C}^{ran}_1\}$ are computed.

\subsubsection{Results}
The simulation results are shown in Fig. \ref{fig:exp2} and note that when $\sigma_{loc}=0.01m,0.1m$ in both 2D and 3D scenarios, $\min\{tr\{\bb{C}^s_1\}\}-tr\{\bb{C}^o_1\}$ is larger than zero and extremely small. There are two observations from Fig. \ref{fig:exp2} and the note, and these observations are consistent in both 2D and 3D cases. Firstly, $tr\{\bb{C}^o_1\}$ is the smallest over all cases, which means OSP-SLN is equivalent to OSP-WSLN. The second one is when sensor location noises are not large ($\sigma_{loc}=0.01m,0.1m$), every $tr\{\bb{C}^s_1\}$ is close enough to $tr\{\bb{C}^o_1\}$ and is less than all of $tr\{\bb{C}^{ran}_1\}$, which supports the second conclusion in this section.


\subsection{Improvement for Source Localization by OSP-SLN}
\subsubsection{Set up}
The superiority of OSP-SLN in sound source localization is verified in this section. The OSP-SLN in 2D space and the OSP-SLN in 3D space are a uniform angular array (UAA) and a platonic solid respectively, which has been demonstrated by the equivalence between OSP-SLN and OSP-WSLN. The configuration setting of UAA and platonic solid are the same as that in Section IV.B. We set $\sigma_t=0.1ms$ and $\sigma_{loc}=0.01m, 0.1m, 1m$. In each $\sigma_{loc}$, we perform 200 Monte-Carlo simulations. Each simulation generates measurements, selects initial values randomly, and localizes the sound source. To be specific, the MSE of the source location following the OSP-SLN strategy (MSE-OSP-SLN) and the MSE of the source location by placing sensors randomly (MSE-RAN) are computed. The formulation of optimization for sound source localization with sensor location errors is
\begin{equation}
    \label{NLS-Loc}
    \min_{\boldsymbol{\theta}} (\bb{g}(\boldsymbol{\theta})-\bb{z})^T\bb{\Sigma}^{-1}(\bb{g}(\boldsymbol{\theta})-\bb{z}),
\end{equation}
which is solved by the Gauss-Newton (GN) method.

\begin{figure}[htbp]
\centerline{\includegraphics[width = 0.97\linewidth]{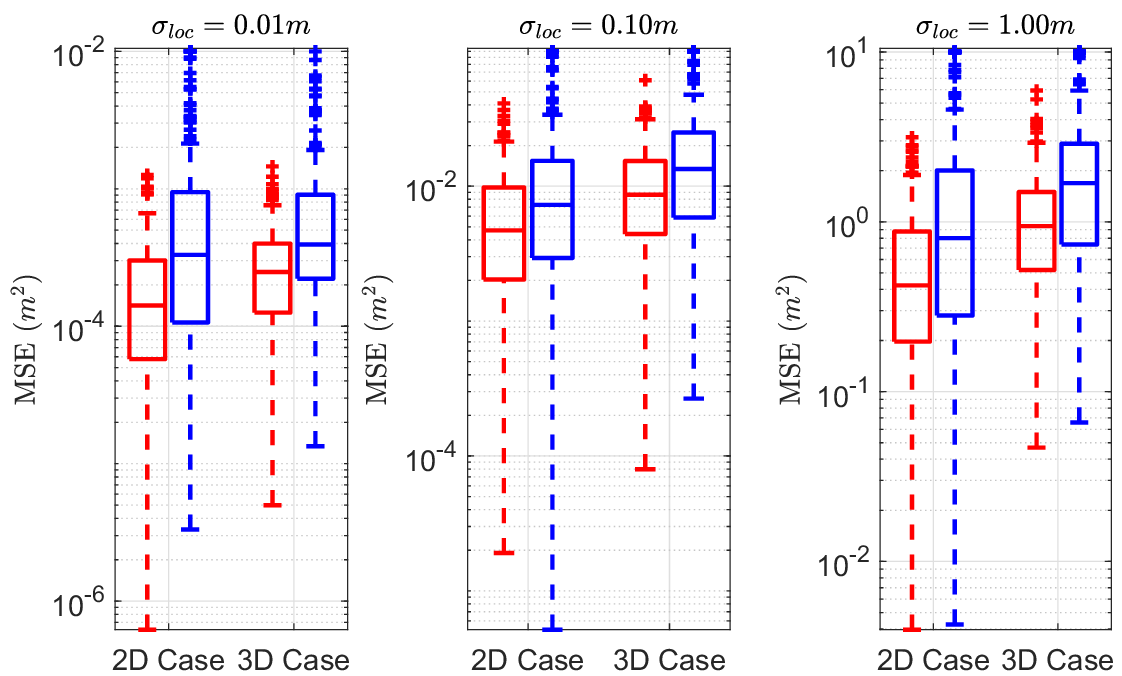}}
\caption{Simulation results of MSE-OSP-SLN (red) and MSE-RAN (blue) in 2D/3D space under three levels of sensor location noises ($\sigma_{loc}=0.01m, 0.1m, 1m$).}
\label{fig:exp3}
\end{figure}

\subsubsection{Results}
The simulation results are shown in Fig. \ref{fig:exp3}. We can observe that the performance of MSE-OSP-SLN is better than that of MSE-RAN in both 2D/3D space under any level of sensor location noise, which means source localization accuracy is improved by the OSP-SLN strategy significantly.

\section{Conclusions and Future Works}
In this paper, we study the optimal sensor placement strategy for TDOA-based source localization under sensor location noises, called OSP-SLN. Under the assumptions of the full set of TDOA and near-field, OSP-SLN is shown to be equivalent to OSP-WSLN from a concise equity Eq. (\ref{eq:linearEq}) that describes the linear relationship between the trace of CRB without sensor location errors and the trace of CRB with the errors. The other problem is how sensor location errors influence the performance of OSP-SLN policy and the answer is obtained in simulation. Extensive simulations validate both equality and equivalence and demonstrate that no large sensor location errors bring an ignorable impact on the performance quantified by the trace of CRB. Further, simulations show that deploying the OSP-SLN policy improves source localization accuracy remarkably. 

In the future, we will study theoretically the impact of sensor position errors on the trace of CRB for the OSP-SLN by modeling $\bb{x}$ as random variables. Also, OSP-SLN strategies will be explored based on other optimization criteria and other measurements.

\bibliographystyle{IEEEtran}
\bibliography{ref}
\end{document}